\documentclass[fleqn,10pt]{wlscirep}

\usepackage{fourier}

\title{Nanoscale neural network using non-linear spin-wave interference}

\author[1]{\'{A}d\'{a}m Papp}
\author[2]{Wolfgang Porod}
\author[1,*]{Gyorgy Csaba}
\affil[1]{P\'{a}zm\'{a}ny P\'{e}ter Catholic University, Faculty of Information Technology and Bionics, Budapest, Hungary}
\affil[2]{Center for Nano Science and Technology University of Notre Dame (ND\textit{nano}), Notre Dame, IN, USA}

\affil[*]{csaba.gyorgy@itk.ppke.hu}

\begin{abstract}
We demonstrate the design of a neural network, where all neuromorphic computing functions, including signal routing and nonlinear activation are performed by spin-wave propagation and interference. Weights and interconnections of the network are realized by a magnetic field pattern that is applied on the spin-wave propagating substrate and scatters the spin waves. The interference of the scattered waves creates a mapping between the wave sources and detectors. Training the neural network is equivalent to finding the field pattern that realizes the desired input-output mapping. A custom-built micromagnetic solver, based on the Pytorch machine learning framework, is used to inverse-design the scatterer. We show that the behavior of spin waves transitions from linear to nonlinear interference at high intensities and that its computational power greatly increases in the nonlinear regime.  We envision small-scale, compact and low-power neural networks that perform their entire function in the spin-wave domain.
\end{abstract}
\begin{document}

\flushbottom
\maketitle
\thispagestyle{empty}

\section{Introduction}
\label{Intro}

The interest in neuromorphic computing hardware skyrocketed in recent years, for two main reasons. It was realized long ago that digital systems (let they be CPUs or GPUs) are rather inefficient for such inherently analog tasks. A more recent development is that traditional, MOS-transistor based devices turned out to have a strong staying power for Boolean, digital logic - which has driven the research of emerging nanoelectric devices towards neuromorphic, analog problems. These are the application areas where emerging devices have the potential to show substantial benefits over MOS switches \cite{ref:grollier}.


A central challenge of the research on neuromorphic devices is that most computing models require highly interconnected systems, i.e. artificial neurons with a large number of connections, often all-to-all connections. Stand-alone neuronal units has little utility - there should always be an effective way to interconnect those devices to computing systems. This is where wave-based computing concepts show their strengths: if the computing device is realized in a  wave-propagating substrate, then interference patterns realize an all-to-all interconnection between points of this substrate. 

The power of wave-based computing have long been harnessed in optical computing and the high interconnectivity is a major selling point for most optical (holographic, interference-based) devices.  It is, however, clear that while \emph{linear} interference is excellent for high interconnections, its computing power is fairly limited. Linear interference is sufficient only for signal processing tasks: general-purpose computing and all variants of neuromorphic computing require some sort of nonlinearity. In optical computing, implementing nonlinearities requires high optical intensities and nonlinearities are often implemented separately from the linear scatterer that provides the interconnections. Other types of waves may implement nonlinear functions in a more natural way. In the present paper we show that spin waves provide both high interconnections and the nonlinearities required for neuromorphic computing. 

Spin waves (also referred to as magnons) are wave-like, collective excitations of a spin ensemble. Here we restrict ourselves to spin waves propagating in ferro-, and ferrimagnetic thin films. Spin-wave behavior is approximately linear at low amplitudes, but nonlinearities become significant at moderate intensities. Unlike photons, magnons interact with each other, which is a requirement for non-trivial computation. Spin waves show many similarities to electromagnetic waves and preserve many benefits of optics, for example, they can maintain long coherence length even at room temperature \cite{ref:grundler}. Spin waves exist down to sub-100 nm wavelengths at microwave frequencies and they are suitable for integration with electronic components \cite{ref:perspectives}. 

High connectivity and built-in nonlinearity make spin waves an ideal choice for neuromorphic computing, in theory. However, in order to actually use spin waves for useful computing tasks, an inverse problem must be solved: one must find a scatterer configuration that yields to a certain input / output relation via the formation of an interference pattern. This is in general a daunting task due to the complexity of nonlinear wave propagation. 

Very recently, Hughes et. al. \cite{ref:stanford} presented a theoretical framework for implementing a recurrent neural network (RNN) in a medium described by a nonlinear wave equation. Specifically, it was shown that if a substrate is described by the nonlinear wave equation and this substrate is excited and probed at given points, then the equations that give the wave dynamics between the prescribed points map to an RNN. In their work the nonlinearity of the medium is modeled by a spatially varying and intensity-dependent wave propagation speed. Training of the neural network is implemented by adjusting the spatially dependent wave propagation speed by gradient-based computational learning. 

The work of Hughes et al. is an original, fresh approach to wave-based computing, but leaves crucial questions unanswered. It is admitted that numerical simulations with the computational learning machine do not fully support the premise of the paper, as the RNN-equivalent nonlinear structure shows similar performance to what is achievable by linear propagation. Thus, it is not proved that the presented structure can indeed exploit nonlinear waves to achieve better performance in problems beyond linear signal processing. Furthermore, it is not elaborated, how the form of nonlinearity assumed in the paper can be realized by a physical system, albeit a few hints are provided for optical implementations. In the present paper, we use the work of Hughes et al.\cite{ref:stanford} as a starting point, but we study an experimentally realizable magnetic system and model it with full micromagnetic simulations that can precisely describe experimental scenarios. We employ a specific physical system and program it to do true neuromorphic functions. The device is a magnetic thin film, with a spatially non-uniform magnetic field acting on it. A custom micromagnetic solver based on a machine-learning framework, Pytorch \cite{ref:pytorch}, is used to design a magnetic field distribution that steers (scatters) spin waves to achieve the desired function. We named our micromagnetic design engine Spintorch.

For small-amplitude excitations, Spintorch solves an inverse problem for the linear wave equation - it designs a magnetic field distribution that performs a desired linear operation (such as matrix multiplication, convolution, pattern matching, spectral analysis, matched filtering) as we will show in Sec. \ref{sec:training}. The algorithm has great utility already in this regime, as it automatizes the design of spin-wave based RF signal processors. 

Higher-amplitude spin waves, with a precession angle above few degrees, show nonlinear behavior and Spintorch -- the exact same computational learning engine -- can be used to design a nonlinear interference  device. This device is functionally equivalent to the RNN of Hughes et al.\cite{ref:stanford}, and in Sec. \ref{sec:nonlinear} we will show how the introduction of nonlinearity increases the computational ability of the device. The spin-wave scatterer becomes a true neural network, exploiting nonlinearity to exceed the performance of linear classifiers.

It is noted, that our manuscript is submitted simultaneously with \cite{ref:andrii_new}, in which the Authors use the inverse-design magnonics to create arbitrary linear, nonlinear and nonreciprocal devices.

\section{Design of spin-wave scatterers by computational learning}
\label{sec:training}


A spin-waves scatterer is a magnetic thin-film  with spatially non-uniform magnetic field acting on it: this magnetic field distribution locally changes the dispersion relation of the wave \cite{ref:apapp}, scatters (steers) the spin waves, creating an interference pattern. For sake of concreteness we assumed that the wave source is a microwave coplanar waveguide (CPW). The output of the spin wave scatterer is the spin-wave intensity at particular areas, which experimentally could be picked up via antennas on the film surface.

\begin{figure}[!htb]
\centering
\includegraphics[width=6.9in]{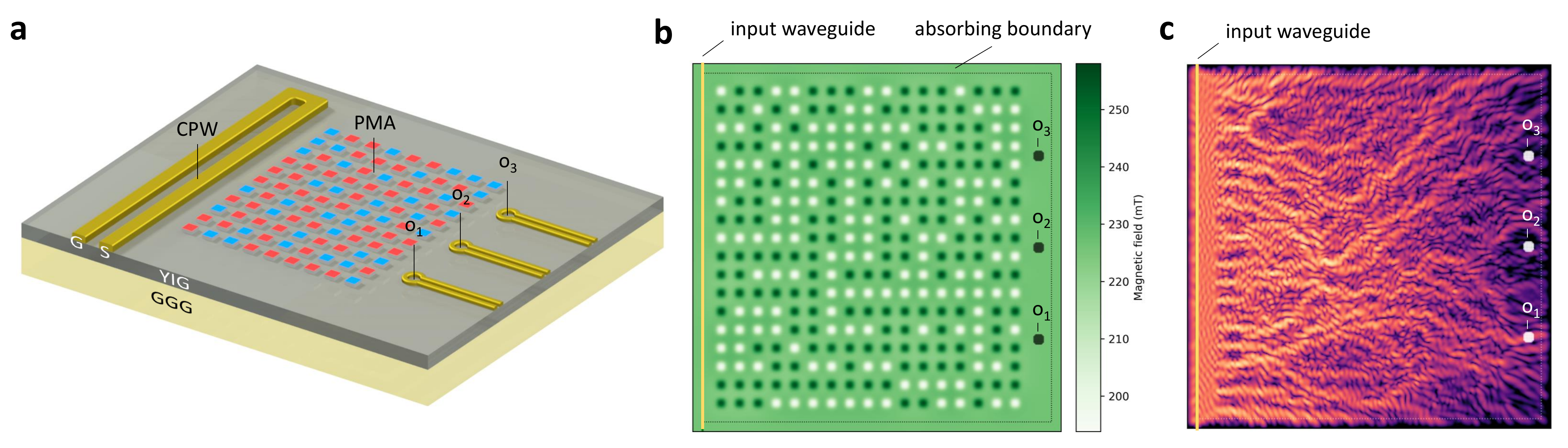}
\caption{Nanomagnet based spin-wave scatterer. \textbf{a}) The schematics of the envisioned computing device. The input signal is applied on the coplanar waveguide (CPW) on the left, and the magnetic state (up/down) of programming magnets on top of the YIG film define the weights. \textbf{b}) The PMA magnets on top of YIG generate a bias-field landscape. The training algorithm finds the binary state of the programming magnets. \textbf{c}) Spin-wave intensity pattern for a particular applied input, which results in a high intensity at $o_1$. The size of the simulation area is 10 µm by 10 µm.} 
\label{fig:fig_punchcard}
\end{figure}

In order to design experimentally realizable field distributions, a specific geometry of field-generating  nanomagnets was assumed, as sketched in Fig. \ref{fig:fig_punchcard}\textbf{a}.  The punchcard-like pattern of up/down pointing nanomagnets sits on top of a low-damping substrate (such as yttrium iron garnet, YIG) and acts as the program for the spin-wave scatterer. The programming nanomagnets assumed to exhibit strong perpendicular magnetic anisotropy (PMA) and their magnetization is not influenced by the spin waves propagating in the layer underneath - see Supplementary for details on the material system.  The physical system is straightforwardly realizable, in fact it is rather similar to the scenarios used in recent experiments, such as in  \cite{ref:grundler}. For some simulations we used a more fine-grained field distribution, see Supplementary for details. 

Spintorch inverse-designs the up/down configuration of the programming magnets in order to realize particular output intensity patterns as a response to an input temporal waveform. The code uses the same gradient-based algorithm that is implemented in Wavetorch \cite{ref:wavetorch,ref:pytorch}, but a GPU-based, custom-built full micromagnetic solver is used to model spin-wave propagation as described in the Supplementary section. Instead of using a (nonlinear) wave equation for modeling wave propagation, we solve for the underlying physics by discretizing the modeled region into 25~nm~$\times~$~25~nm~$\times~$~25~nm sized volumes and  solve the Landau-Lifshitz-Gilbert equations \cite{ref:mumax} to calculate the precession of magnetic moments in these computational regions. Most importantly, our micromagnetic solver fully accounts for the demagnetizing field, and thus the change in magnetic field due to the magnetization precession, which is the source of nonlinearity in spin-wave propagation. The micromagnetic solver is fully integrated within the computational engine which performs gradient based optimization of the trainable parameters finding the optimal up/down magnet configuration.

\subsection{Inverse design in the small-amplitude regime}
\label{sec:vowel}

Perhaps the simplest example of inverse design is that of a spectrum analyzer, where the design objective is to focus different spectral components (frequencies) to different spatial locations of the scatterer. In our example we used a 10 µm $\times$ 10 µm scatterer to separate 3 GHz, 3.5 GHz and 4 GHz components of the time-domain signal applied on the waveguide. The outputs  are 300 nm diameter areas and time-integrated wave intensity over these areas is defined as the output variable.

\begin{figure}[!ht]
\centering
\includegraphics[width=6.9in]{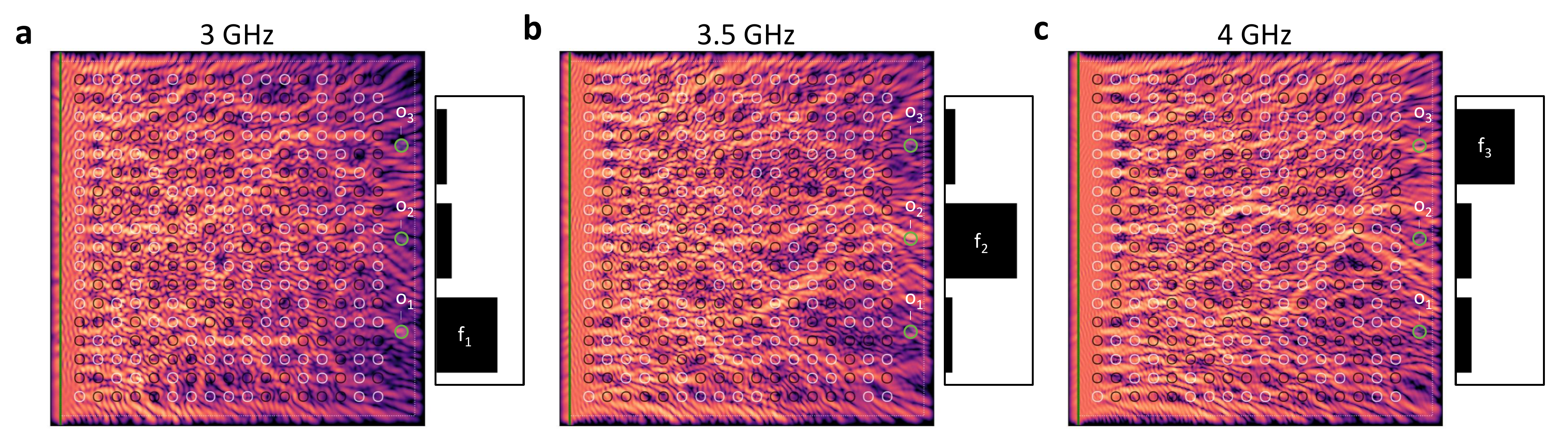}
\caption{Frequency separation by training. \textbf{a}-\textbf{c}) The scatterer was trained to direct frequency components $f_1$~=~3~GHz, $f_2$~=~3.5~GHz, and $f_3$~=~4~GHz to the corresponding outputs denoted by $o_1$, $o_2$, $o_3$. The bar charts indicate time-integrated intensities measured at the outputs (green circles). The colormaps show time-integrated intensity of spin waves at t~=~30~ns. Black/white circles are contours of the out-of-plane component of the magnetic field indicating the state of the magnets on top of the YIG film (same in all cases \textbf{a}-\textbf{c}). The size of the simulation area is 10 µm by 10 µm.} 
\label{fig:fig_spectrum}
\end{figure}

The computational learning engine converges to a high-quality design in about 30 training epochs. Here we used small-amplitude spin waves for the training: for  precession angles not exceeding a few degrees (excitation fields in the mT range), the computational learning algorithm finds the same solution regardless of the amplitude. The snapshots of Fig. \ref{fig:fig_spectrum} show the spin-wave intensity for the the three frequencies and show that the device performs the required function. The punch-card program that is found by the learning engine is non-intuitive and does not resemble spectrum analyzer designs that were constructed from optical analogies \cite{ref:rowland}. The field pattern however, makes a similar impression to refractive index patterns in photonic metamaterial devices \cite{ref:inverse01,ref:inverse02, ref:engheta}. The converged scattering pattern also depends on initial conditions that are given to the computational learning engine. The designs, however, all appear to be robust: we verified that switching errors in the magnet states (which are unavoidable in an experimentally realized device) do not affect the performance significantly in most cases. 

We would like to point out that the selectivity of the spectrum analyzer design is limited by the relatively small degrees of freedom provided by the approximately 300 binary values (i.e. the magnets). To scale up the simulations to include more magnets, significantly more computing resources would be needed. Instead, in the following examples we used external magnetic field values as training parameters directly without simulating magnets on top. This increases the degrees of freedom to approximately 1600 continuous variables, resulting in much better performance at the same computational expense. 

The automatized design of linear signal processors alone is an important results and opens many potential application for spin-wave-based devices.  Just as photonic metamaterials have much smaller footprint than classical optics devices (such as a $4f$ correlator), the above-designed scatterer (spin-wave metamaterial) has the same advantages over designs based on classical optics (such as \cite{ref:lens, ref:rowland}).

\subsection{Vowel recognition in the linear and nonlinear regime}
\label{sec:vowel2}

For the computational engine it makes no difference whether the scatterer needs to focus 'pure' frequencies to the output points or it has to identify a certain  spectral pattern. We tested this by running a vowel recognition example using the vowel samples available in the Wavetorch package \cite{ref:wavetorch}\cite{ref:stanford}. The waveforms of the vowels were scaled up to microwave frequencies, in such a way that the frequency components with significant energy content on the input waveguide launch propagating waves with wavelengths compatible with the scatterer. The scatterer structure was trained to maximize the spin-wave intensity at one of the three output points, which correspond to the recognized vowels. We used two or four samples of each vowel as a training set. The rest of the 44 samples for each vowel was used as test set.

\begin{figure}[!ht]
\centering
\includegraphics[width=6.5in]{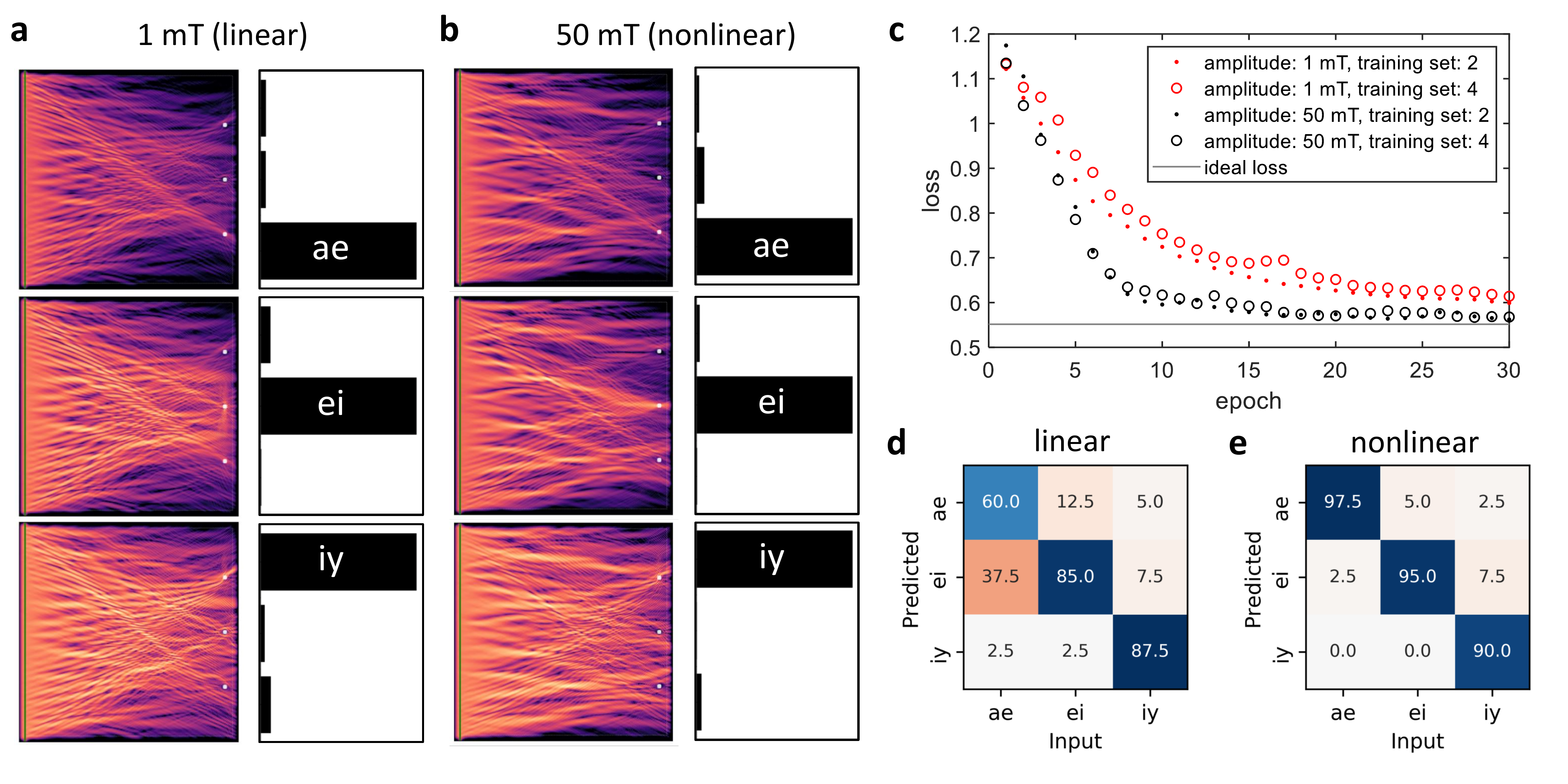}
\caption{Using the spin-wave scatterer for vowel recognition. \textbf{a-b}) Wave intensity patterns, formed in response to the time-domain excitations (vowels). The scatterer was trained to focus waves to the corresponding outputs. The bar charts show the intensity at the output locations (normalized). The linear regime \textbf{a} (1~mT excitation field) and the nonlinear regime \textbf{b} (50~mT excitation field) performs comparably well on the training data (slight improvement in case of nonlinear waves). \textbf{c}) Cross-entropy loss decreases during the training, indicating learning. After 30 epochs (training steps), the nonlinear cases achieve better performance compared to the linear case. Note that a nonzero loss value corresponds to the perfect response, indicated by solid line. \textbf{d-e}) Confusion matrices over the testing data set (120 vowel samples).} 
\label{fig:fig_vowel}
\end{figure}

Some results on the training samples can be seen in Fig. \ref{fig:fig_vowel}\textbf{a}. In 30 training epochs the system was able learn to distinguish the vowels ‘ae’, ‘ei’, and ‘iy’, directing the waves toward the correct outputs.

For comparison, we repeated the simulations with increased excitation fields (nonlinear regime, see Fig. \ref{fig:fig_vowel}\textbf{b}). On the training dataset the difference is not very significant, although in Fig. \ref{fig:fig_vowel}\textbf{c} it is clearly visible that the nonlinear operation achieved better performance and also the convergence is faster. The quality of the vowel recognition operation is also compared using confusion matrices in case of the testing dataset. For three vowels these are $3 \times 3$ matrices, where the rows correspond to the predicted output, the  columns to the applied input and the $c_{ij}$ matrix elements give the percentage of cases where vowel $i$ is identified for $j$ vowel as input. For perfect recognition the confusion matrix is diagonal with $100 \%$ at the $c_{ii}$ elements. 

Confusion matrices on the training data set – not shown here – were perfectly diagonal for all amplitudes, which is not surprising with these small training sets and the comparably large internal complexity of the scatterer. But a striking difference appears between linear and nonlinear performance in case the larger, 44-element data set is used for testing. Figure  \ref{fig:fig_vowel}$d$ shows the confusion matrices for this test scenario. Confusion matrices are significantly more diagonal in the non-linear case. The linear device performed significantly worse in learning to recognize unseen vowels. The nonlinear device, however, performed relatively well even when trained only on two samples (not shown here) and its accuracy further improved when using four samples for training. In the latter case it misidentified only 7 out of 120 vowels. 

The confusion matrices in this test scenario characterize the generalization (extrapolation) ability of the network. Based on a very small (2 or 4 vowel waveform) learning set, the network had to recognize and classify vowels that it had not seen before. Linear scatterers can not excel in this job - they match the distinctive spectral features of learned samples, but their ability to generalize from learned data is very limited. The nonlinear scatterer appears to behave as a true neural network, which performs nonlinear classification and generalizes (extrapolates) from the training data. We believe that our simulation data may also verify the hypothesis of \cite{ref:stanford}, that nonlinear wave interference acts as an RNN.

\subsection{Linear vs. nonlinear interference in the scattering block}
\label{sec:nonlinear}

Performing successful vowel recognition does not necessarily require a neural network and satisfactory  results can be obtained by linear classifiers, as shown in the above example. Using the vowel recognition example, it is not at all straightforward to identify what benefits could possibly come from using a neural network like behavior.

\begin{figure}[!ht]
\centering
\includegraphics[width=6.5in]{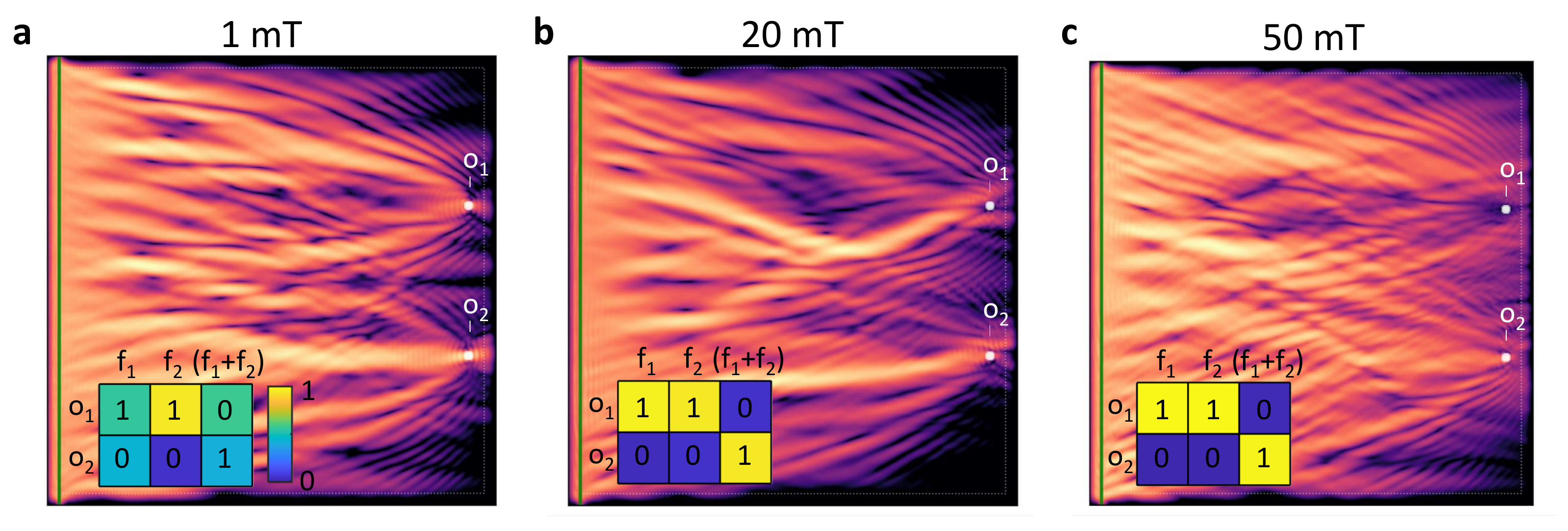}
\caption{ A simple example of a problem that is not solvable by a linear system. Input is encoded in two frequencies ($f_1$~=~3 GHz and $f_2$~=~4 GHz), and the training function is listed in the inset tables (expected results indicated by numbers, output data is shown in color). \textbf{a}) In the linear case (1 mT excitation field) application of simultaneous frequencies results in both $o_1$ and $o_2$ high (incorrect training). \textbf{b}-\textbf{c}) In the nonlinear cases the wave is focused at $o_2$, but $o_1$ is avoided (correct operation). In case of 50 mT excitation the distinction is even stronger. The colormap shows integrated wave intensity. The size of the simulation area is 10 µm by 10 µm. } 
\label{fig:fig_nonlin}
\end{figure}

However, a fairly simple example can show the computational limitations of linear interference, where the superposition principle always holds. In the following example (see Fig. \ref{fig:fig_nonlin}) the training goals  were to 
\begin{enumerate}[label=(\Alph*)]
\item focus waves on output $o_1$ at 3 GHz input frequency,
\item focus waves on output $o_1$ at 4 GHz input frequency, and
\item focus waves on output $o_2$ if and only if 3 GHz and 4 GHz are simultaneously present.
\end{enumerate}
 Clearly, condition (C) is inconsistent with the superposition of (A) and (B).

We used excitation amplitudes in the linear (1 mT) and soft-nonlinear (20 mT and 50 mT) regimes and run the training for 30 epochs in every case. The resulting spin-wave intensity snapshots are shown in Fig. \ref{fig:fig_nonlin}. It is clearly visible that the results of the training are different in case of different amplitudes. The paths traveled by the waves are completely different in the three cases. It is also clear from the snapshots that the linear case failed to focus on $o_2$, while the nonlinear cases were clearly focusing to the bottom output ($o_2$) avoiding $o_1$. 

As expected, the linear case could not provide the desired outcome: the output of the two-frequency case is a linear combination of the outputs observed with single-frequency excitations. On the contrary, the operation in the nonlinear regime achieved  good results, with the highest amplitude excitation giving the best outcome. Quantitatively, the loss function, which quantifies the quality of the computational learning, yields the same conclusion: The linear case did not show a convergence over the 30 epochs, while the nonlinear cases converged to an optimal loss value. The highest excitation amplitude achieved lower loss at the end of the training, and its convergence was also faster.

This elementary example demonstrates a crucial difference between the computational power of linear and nonlinear spin waves - and this difference is expected to manifest itself for more complex operations, such as the high-amplitude vowel recognition example in Sec \ref{sec:vowel2}. It also serves as a proof that our computation engine is able to exploit the nonlinearity of spin waves.

\section{Benchmarks: computing power of the spin-wave substrate}
\label{sec:benchmark}

Nowadays complex neuromorphic computing pipelines are implemented on CPUs and GPUs, which posses extreme computing power, but have poor energy efficiency for analog neuromorphic tasks: computing steps are implemented on digital (often floating point) arithmetic and each of those steps consume energy in the $E=10^{-11}$ J range. A typical processing computing primitive (such as a single convolution with smaller precision in Convolutional Neural Network) consumes roughly the same amount.

Spin waves in nanoscale magnetic structures carry very little energy: the total magnetic energy stored in the patterns of say Fig. \ref{fig:fig_nonlin}\textbf{c} is about $E \approx 1000$ eV, $E \approx 10^{-16}$ J. Patterns in the linear regime hold orders of magnitude less energy, in the few-eV range. The stored  energy can serve as a first estimate of the energy that is dissipated in the magnetic domain in each neural computation step. The time it takes for the interference pattern to build up is in the order of $t=10$ ns -- so the spin-wave scatterer simultaneously achieves low power and high speed. These are stellar numbers when compared to the above-mentioned energy of approximate convolution or floating point operation, indicating great potential to the spin wave based processor. Neurons and synapses based on other emerging devices \cite{ref:nikonov} also consume significantly more energy than the stated $E \approx 10^{-16}$ and they do it at slower computational times. The neural operation done by the scattering block in the nonlinear regime is also considerably more complex than a convolution or what is performed by the synapses and neurons in \cite{ref:nikonov}.

The energy dissipated in the magnetic domain is just a lower bound for consumption: the spin-wave scatterer is most likely used as a hardware accelerator in electrical circuitry -- and in that case the net energy efficiency of spin-wave-based computing block is dominated by the magneto-electric transducer. More specifically, picking up magnetic oscillations from sub-square-micrometer areas will induce less then a microvolt voltage in the transducer antenna, possibly even less then that. Amplifying such small and high-frequency signals requires significant microwave circuitry, which consumes at least 10 mW of power \cite{ref:egel}. Assuming a GHz date rate, this gives $E=10^{-11}$ J per output point. Transduction on the input side (creation of spin waves) is less of a problem as can be done by acceptable efficiency using coplanar waveguides and a single waveguide can excite a larger number of scatterers. 

The net power efficiency of the spin-wave scatterer is comparable to that of electronic implementation for a simple operation (i.e. a convolution). If large internal complexity can be reached in the scatterer with a single or very few inputs, then the spin-wave scatterer potentially leads to several orders of magnitude performance gain compared to electronic implementations.

It is worth noting that optical reservoir computing \cite{ref:reservoir01,ref:reservoir02} -- another promising hardware for accelerating neural computations -- consumes in the order of $E=10^{-11}$ to $E=10^{-12}$, which is comparable to a small spin-wave scatterer with I/O, with a significantly larger device footprint. Strictly linear operations in optics may be performed with much higher energy efficiency (due to the more straightforward scalability of optical systems \cite{ref:tensor}), but such systems require several additional components for general-purpose computation.

\section{Conclusions}
\label{sec:conclusion}

Spin waves are a leading candidate for non-electrical information processing and magnonic devices have been designed for many different purposes, such as Boolean logic gates \cite{ref:chumak}, and signal processors \cite{ref:rowland}. In many cases, magnonic computers are derived from photonic computing devices and most often classical photonics is used as an inspiration, with lenses, mirrors, interferometers designed in the spin-wave domain.

Our work advances the state of art of magnonic computing devices on two fronts. Firstly, we demonstrated that the computational tools developed for the inverse-design of photonic metamaterials (a.k.a. photonic inverse design) can be applied in the spin-wave domain: convolvers, spectrum analyzers, matched filters and possibly a large variety of RF signal processing devices can be designed in a fully automatic way. Spin waves, unlike electromagnetic waves, seamlessly transition to a nonlinear regime at higher excitation amplitudes. Apparently, the computational design algorithm operates just as well if the underlying wave propagation is nonlinear wave, and designs devices based on nonlinear interference. The second and perhaps most important result of our work is that the capabilities of such-designed nonlinear interference devices go way beyond linear signal processing, they are likely equivalent to recurrent neural networks. The device   realizes all the interconnections, weighted sums and the nonlinearities in a single magnetic film.

Wave-based general-purpose computing -- and more generally, computing in a material substrate by the laws of physics -- is a longtime dream of the emerging computing community \cite{ref:stepney, ref:porod, ref:csaba, ref:solitons}. Possibly, spin-wave-based nonlinear processors bring closer to the fulfillment of this vision.

\bibliography{bib}

\section*{Acknowledgements}
The authors are grateful for fruitful discussions and encouragement from Markus Becherer (TUM), Andrii Chumak, Qi Wang (University of Vienna) and Philipp Pirro (TU Kaiserslautern). We are also grateful for the team of the DARPA NAC (Nature as Computer) program, especially to Dr. Jiangying Zhou (DARPA) for her professional project management and expert advice to drive our the work. This research was in part financially suported by the DARPA NAC  program. Adam Papp received funding from the postdoctoral grant (PPD-2019) of the Hungarian Academy of Sciences.

\section*{Author contributions statement}

G. C. and  W.P.  conceived the original idea, \'{A}.P. designed the computational engine, performed the micromagnetic simulations and explored the role of nonlinearities. \'{A}.P. W. P. and G.C. wrote the manuscript. All authors discussed the results and reviewed the manuscript. 

\section*{Additional information}
\textbf{Competing financial interests:} 
The authors declare no competing financial interests.

\begin{small}

\section{Supplementary materials / methods}

\subsection{Implementation of the computational learning engine}

Spintorch is a modified version of Wavetorch \cite{ref:wavetorch}, in which we implemented a full micromagnetic solver to precisely model spin-wave behavior. The numerical engine for inverse design is built on the popular and open-source machine learning framework, Pytorch \cite{ref:pytorch}. An important feature of Pytorch is the automatic gradient calculation, which allows automatic backpropagation throughout complicated multilayered computational flows. In our system this means gradient calculation can be performed backwards in time throughout the whole wave propagation. This allows us to perform gradient based optimization of the trainable parameters, e.g. the applied magnetic-field distribution. Pytorch also provides a number of optimizers, loss functions, and data loading modules, so we did not need to implement these from scratch. Pytorch modules can run on CPU or GPU (using CUDA), without any device-specific coding on the user side. 

In order to exploit the automatic gradient calculation feature of Pytorch, custom modules must use the internal methods for implementing the forward path of the system. This way the backward method is automatically generated on the fly by building a computational graph and saving the required intermediate results. Thus, readily available micromagnetic solvers (such as OOMMF or  mumax3 \cite{ref:mumax}) cannot be integrated in Pytorch because these do not build a computational graph and do not save intermediate results for backpropagation. 

\subsection{Modeling the computing substrate}

The dynamics of the media are described by the Landau-Lifsitz-Gilbert (LLG) equation and takes into account all relevant physical interactions in the micromagnetic model. Elementary magnetic moments are represented by three-dimensional vectors, and we use a finite difference discretization with a rectangular grid.  The dynamics of magnetic moments depend on the torque exerted on them by the effective magnetic field, which is a sum of several field components. Most importantly, it includes a (space-, and time-dependent) external field, the dipole fields of other magnetic moments, and the exchange interaction between neighboring electrons. The dipole interaction is a long-range effect, thus, it is the most computationally expensive part of the calculation. We used FFT based acceleration for calculating the solution of the Poisson equation (i.e. determine the dipole fields), for which the GPU-accelerated FFT module of Pytorch enabled effective implementation. The exchange field is calculated only between nearest neighbors, as exchange field is local. Exchange-field calculation is implemented using a convolution with a Laplacian kernel. The time stepping of the differential equation is realized by a classical 4th order Runge-Kutta method. The LLG equation also includes a damping term, which is also implemented in our code, thus realistic attenuation of spin waves is simulated. Damping is also used to realize absorbing boundary conditions, so we could accurately model a few-micron sized, finite region of an extended magnetic film. The micromagnetic model fully accounts for the nonlinearities appearing at higher intensities: these nonlinearities are a direct consequence of the dependence of the demagnetizing field on the local magnetization and the spin-wave amplitude.

We verified our solver by comparing results with the widely used mumax3 solver \cite{ref:mumax}. The high-level use of GPU-based functions and the overhead of automatic gradient computation makes our code less efficient as a general purpose micromagnetic solver, but still, running times are comparably fast and more than 100,000 cells with a few thousand timesteps can be simulated in minutes on a state-of-the-art GPU. This makes it possible to embed the solver into the learning algorithm and train the system with multiple samples and epochs within a few hours or, with a larger training set, days.

\subsection{Magnetic material properties}

YIG is used as a medium for low-damping spin-wave propagation, and arrays of nanomagnets with perpendicular magnetic anisotropy (PMA) provide control over spin waves via their dipole fields. PMA magnets are bistable (magnetization pointing either upwards or downwards) if their size is below the single domain limit (typically less than a few hundred nanometers).

Such a system could provide a reconfigurable means to programming spin-wave-based neural networks, by individual switching of the nanomagnets. This implementation of the scatterer shows many benefits over a lithographically patterned (hardwired) scatterer. In our model we included the calculation of realistic dipole fields of the nanomagnet arrays, which works for any configuration. 


The chosen material system and geometry is just one of many possible choices. Metallic ferromagnets could have been used in place of the YIG film - these have higher damping (shorter propagation length), but easier to integrate and access electrically. Also instead of the stray-field programming, lithographically defined patterns (lithography followed by etching) could have defined the function of the scatterer. Fine-grained tuning of YIG magnetic properties can be achieved by FIB irradiation of a YIG film \cite{ref:fib,ref:fib2} that continuously changes magnetic parameters as a function of the local dose. We expect that our computational engine can be used with similar effectiveness when film magnetic parameters are adjusted by training, instead of designing the applied field pattern as we have done in this work.

\subsection{Linear spin-wave scatterer as a perceptron layer}
\label{sec:perceptron}

Here we would like to show how a spin-wave scatterer block can represent a single layer of a neural network (perceptron layer). A perceptron layer can be described mathematically as a linear transformation (vector-matrix multiplication) followed by a nonlinear activation function: \(\textbf{y} = \sigma\left(\textbf{Wx}\right)\), where \(\textbf{x}\) is a vector of length \(n\) representing the input, \(\textbf{W}\) is an \(m\times n\) matrix that contains the trainable weights of the perceptron layer, and \(\sigma\) is an activation function applied on every output channel (in the simplest case a threshold function). Regarding its functionality, a perceptron performs a linear classification, so a layer of perceptrons performs \(m\) different linear classifications.  The linear transformation (\(\textbf{W}\)) can be performed by a spin-wave scatterer block, as depicted in Fig. \ref{fig:scatterer}. If the amplitude of the spin-waves is sufficiently small, the wave propagation can be described by the linear wave equation. Input signals routed to input antennas generate spin waves with corresponding amplitudes and phases. Waves travel through a region where the effective refractive index is spatially varying according to the program (the desired linear transformation). The wave intensity from every input will be distributed among the outputs by the scatterer map (some losses may also occur). Since the wave propagation is assumed to be linear, the activation function has to be implemented in the readout circuitry. 

\begin{figure}[!ht]
\centering
\includegraphics[width=5in]{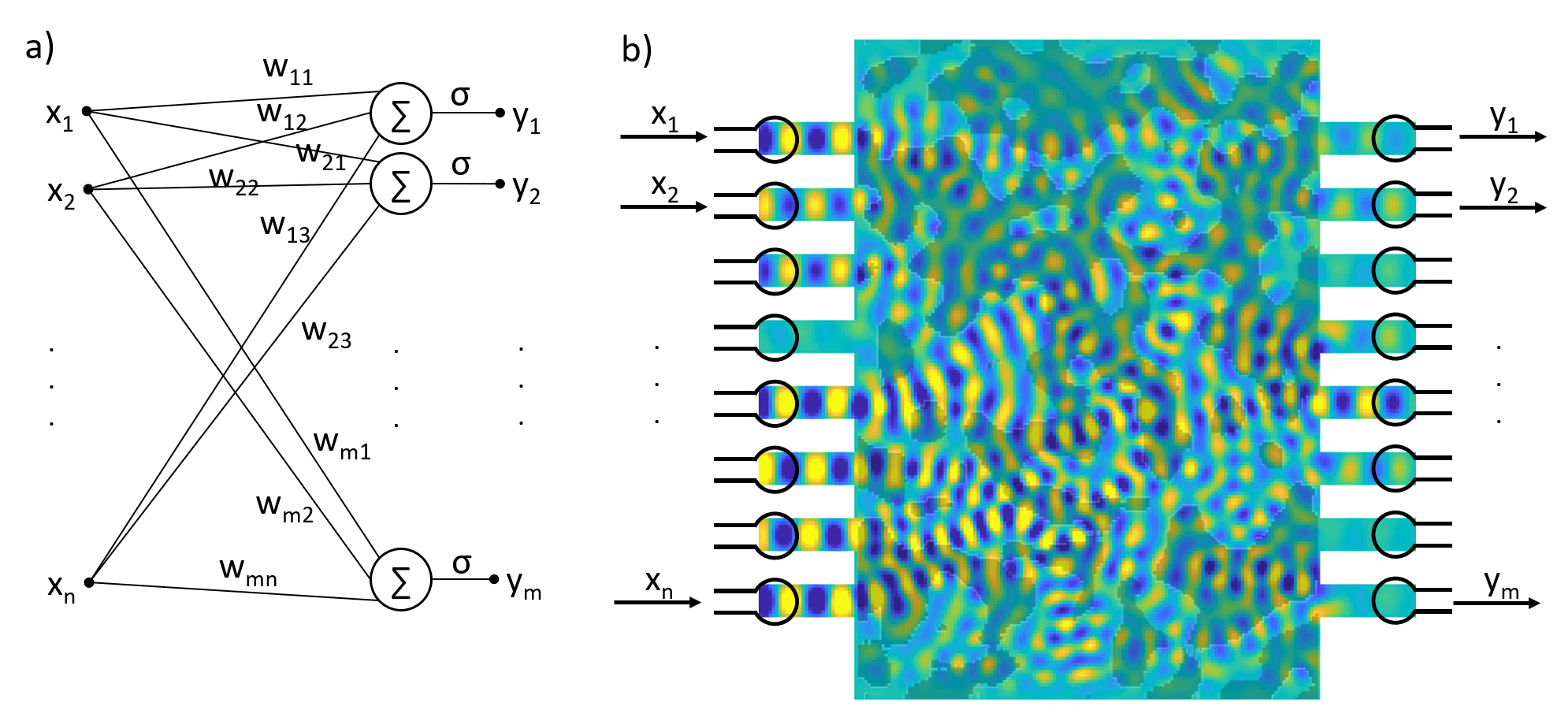}
\caption{a) Perceptron layer b) Spin-wave scatterer} 
\label{fig:scatterer}
\end{figure}

The matrix representation of a given scatterer map can be constructed using the superposition principle: exciting the inputs one at a time with unit amplitude (base vectors) and recording the outputs will produce the columns of the equivalent matrix. The inverse problem is, however, more cumbersome to solve in general. One possible approach is the machine learning method described by Hughes et al. \cite{ref:stanford}, which is directly applicable to any system that obeys the linear wave equation, or can be modified for nonlinear equations. 

Such a device, apart from dynamic range and scaling limitations, can in principle realize any perceptron layer. But the computing capabilities of a single layer are limited to linear classification, and even some relatively simple operations (such as an XOR gate) is impossible to realize using this device. To overcome such limitations one could create a multilayer neural network using such devices sequentially, but any advantages that come from the low-power operation and compactness of the spin-wave scatterer would be overshadowed by the required readout circuitry. Any approach to exploit the benefits of the highly interconnected nature of wave interference should minimize the number of input-output conversions. Thus, we investigated the feasibility of exploiting the nonlinearity of spin waves, which would allow implementing a multilayer neural network (or a recurrent neural network) within a single scattering block.

\end{small}


\end{document}